# On the axisymmetric post-buckling of a thin elastic circular plate under uniform edge pressure

Milan Batista

University of Ljubljana, Faculty of Maritime Studies and Transport, Slovenia

milan.batista@fpp.uni-lj.si

**Abstract**

In this paper, we revisit a post-buckling of an elastic circular plate subject to uniform pressure on its edge. The governing von Karman equations are solved by power series method and numerical colocation method. An alternative method for calculation of coefficients of the series solution is described. It is shown that the power series solution is useful only for the first buckling mode while collocation method can handle also higher buckling modes.

## 1 Introduction

In a middle of the last century, the problem of the post-buckling behavior of a uniformly compressed circular plate was one of the challenge problems of mechanics [1]. In essence, the problem is to find a solution of the von Kármán pair of non-linear partial differential equations under given boundary conditions. Initial analytical research on the clamped plate was done by Way [2]. Friedrichs and Stoker [3-5] provide an extensive study of a simply supported plate using perturbation method, power series method and asymptotic, i.e., boundary layer analysis. In this way, they cover the full range of possible edge pressure (or plate thickness). Bodner [6] apply the power series method to the case of a clamped circular plate. Keller et al. [7] and Wolkowisky [8] investigate the existence of higher buckled states of the plate. Keller and Reiss [9] use finite differences approximation of the von Karman equations to solve the problem.

Sherbourne [8] use Runge-Kutta numerical integration for the solution for a simply supported plate. Rao and Ruju [10] apply FEM to calculate the critical pressure of a clamped and for a simply supported circular plate. Quin and Huang [11] and Ye [12] use BEM to calculate post-buckling deformation for a clamped and for a simply supported circular plate. Faris and Abdalla [13] use the Galerkin method to calculate the post-buckling deflection of a clamped circular plate. Williams et all [14] use collocation method to calculate the axisymmetric post-buckling deflection of a clamped circular plate. Secondary asymmetric buckling of the circular plate was investigated by Cheo and Reiss [15], Chien [16] and Wang [17].

In this paper, we will consider the problem once more. In the next section, which follows the works of Friedrichs and Stoker [3-5], we derive the non-dimensional form of the von Kármán equations of the problem. In the third section, we will consider calculation of the buckling pressure. The fourth section is dedicated to the solution of governing equations with the power series method. The next section deals with numerical solution of governing equations by collocation method [18]. The article ends with conclusions.

## 2 Problem formulation

*Basic equations.* We consider of a thin elastic circular plate with radius $R$ and constant thickness $h$ under uniform pressure $p$ at its edge. A radial symmetric buckling of the plate is described by the von Kármán equations [1]. These are two simultaneous equations for the Airy stress function $\phi$ and for the plate deflection $w$, which can be for radial symmetric buckling written as follows [19, 20]

$$\left.\begin{aligned} \frac{1}{r}\frac{d}{dr}\left(r\frac{d}{dr}\left(\frac{1}{r}\frac{d}{dr}\left(r\frac{d\phi}{dr}\right)\right)\right) &= -E\left(\frac{1}{r}\frac{dw}{dr}\frac{d^2w}{dr^2}\right) \\ D\frac{1}{r}\frac{d}{dr}\left(r\frac{d}{dr}\left(\frac{1}{r}\frac{d}{dr}\left(r\frac{dw}{dr}\right)\right)\right) &= h\left(\frac{1}{r}\frac{d^2w}{dr^2}\frac{d\phi}{dr}+\frac{1}{r}\frac{dw}{dr}\frac{d^2\phi}{dr^2}\right) \end{aligned}\right\} \tag{1}$$

Here $0 \le r < R$ is the radial coordinate, $E$ is Young's modulus, $D \equiv h^3E\big/12\left(1-\nu^2\right)$ is the flexural rigidity of the plate and $\nu$ is the Poisson's ratio. With $\phi$ the membrane stress components $\sigma_r$, $\sigma_\theta$, $\tau_{r\theta}$ are [21]

$$\sigma_r = \frac{1}{r}\frac{d\phi}{dr} \qquad \sigma_\theta = \frac{d^2\phi}{dr^2} \qquad \tau_{r\theta} = 0 \tag{2}$$

The stress resultants that is the bending moments $M_r, M_\theta, M_{r\theta}$ and the shear force $Q_r$, are expressed in terms of $w$ in the following way [22]

$$\left.\begin{aligned} M_r = -D\left(\frac{d^2w}{dr^2}+\frac{\nu}{r}\frac{dw}{dr}\right) \quad M_\theta = -D\left(\nu\frac{d^2w}{dr^2}+\frac{1}{r}\frac{dw}{dr}\right) \quad M_{r\theta}=0 \\ Q_r = -D\frac{d}{dr}\left(\frac{d^2w}{dr^2}+\frac{1}{r}\frac{dw}{dr}\right) \end{aligned}\right\} \tag{3}$$

*Non-dimensional equations.* We introduce the following non-dimensional variables

$$\frac{r}{R}\to r \quad \frac{w}{w_*}\to w \quad \frac{\phi}{\phi_*}\to\phi \quad \frac{\sigma}{\sigma_*}\to\sigma \quad \frac{p}{\sigma_*}\to p \quad \frac{M}{M_*}\to M \quad \frac{Q}{Q_*}\to Q \tag{4}$$

where the scale factors are

$$w_* = \eta R \quad \phi_* = \eta^2 R^2 E \quad \sigma_* = \eta^2 E \quad M_* = \eta\frac{D}{R} \quad Q_* = \eta\frac{D}{R^2} \tag{5}$$

and $\eta$ is the slenderness ratio given by

$$\eta = \frac{1}{\sqrt{12\left(1-\nu^2\right)}}\frac{h}{R} \tag{6}$$

With this the governing equations (1) takes on the dimensionless form

$$\left.\begin{aligned} \frac{1}{r}\frac{d}{dr}\left(r\frac{d}{dr}\left(\frac{1}{r}\frac{d}{dr}\left(r\frac{d\phi}{dr}\right)\right)\right) = -\frac{1}{r}\frac{dw}{dr}\frac{d^2w}{dr^2} \\ \frac{1}{r}\frac{d}{dr}\left(r\frac{d}{dr}\left(\frac{1}{r}\frac{d}{dr}\left(r\frac{dw}{dr}\right)\right)\right) = \frac{1}{r}\frac{d^2w}{dr^2}\frac{d\phi}{dr}+\frac{1}{r}\frac{dw}{dr}\frac{d^2\phi}{dr^2} \end{aligned}\right\} \tag{7}$$

where $0 \le r < 1$. The membrane stresses (2) retains their form, while the expressions for stress resultants (3) become

$$M_r = -\left(\frac{d^2w}{dr^2}+\frac{\nu}{r}\frac{dw}{dr}\right) \quad M_\theta = -\left(\nu\frac{d^2w}{dr^2}+\frac{1}{r}\frac{dw}{dr}\right) \quad Q_r = -\frac{d}{dr}\left(\frac{d^2w}{dr^2}+\frac{1}{r}\frac{dw}{dr}\right) \tag{8}$$

From now one we regard all the variables as non-dimensional.

*Boundary conditions.* Equations in system (7) are of fourth order; therefore, we need eight boundary conditions. These are the following.

At the plate edge $r=1$ we have prescribed pressure $p$, deflection and edge support

$$\sigma_r\left(1\right)=-p \tag{9}$$

$$w\left(1\right)=0 \tag{10}$$

$$\left.\begin{aligned} &\frac{dw}{dr}\left(1\right)=0 \quad \left(\text{clamped edge}\right) \\ &M_r\left(1\right)=0 \quad \left(\text{hinged edge}\right) \end{aligned}\right\} \tag{11}$$

At the plate center we require that the values of the membrane stresses (2) and the stress resultants (8) are finite. To fulfill these conditions, it is at first necessary that

$$\frac{d\phi}{dr}\left(0\right)=0 \qquad \frac{dw}{dr}\left(0\right)=0 \tag{12}$$

By applying the L'Hospital rule to (2) and (8) and using (12), we obtain

$$\sigma_r\left(0\right)=\sigma_\theta\left(0\right)=\frac{d^2\phi}{dr^2}\left(0\right) \tag{13}$$

$$M_\theta\left(0\right)=M_r\left(0\right)=-\left(1+\nu\right)\frac{d^2w}{dr^2}\left(0\right) \qquad Q_r\left(0\right)=-\frac{3}{2}\frac{d^3w}{dr^3}\left(0\right) \tag{14}$$

From this it follows that the membrane stresses and the stresses resultants will be finite at $r=0$ when

$$\frac{d^2\phi}{dr^2}\left(0\right)\neq\infty\,, \quad \frac{d^2w}{dr^2}\left(0\right)\neq\infty\,, \quad \frac{d^3w}{dr^3}\left(0\right)\neq\infty \tag{15}$$

With (9),(10),(11),(12) and (15) all eight boundary conditions are established.

We note that by nondimensionalization any solution of the problem contains only the parameters that configure in boundary conditions (9) and (11): that is $p$ and in a case of a simply supported plate also $\nu$ [3, 4, 19].

*First integral.* Following Stoker [3, 5, 19] we are multiplying each of equations (7) by $r$ and integrate them subject to the conditions (12) and (15). In this way we obtain

$$\left.\begin{aligned} &\frac{d^2\Phi}{dr^2}+\frac{3}{r}\frac{d\Phi}{dr}+\frac{1}{2}W^2=0 \\ &\frac{d^2W}{dr^2}+\frac{3}{r}\frac{dW}{dr}-\Phi W=0 \end{aligned}\right\} \tag{16}$$

where we introduce two new functions

$$\Phi \equiv \frac{1}{r}\frac{d\phi}{dr} \qquad W \equiv \frac{1}{r}\frac{dw}{dr} \tag{17}$$

Once $W$ is determined, we obtain $w$ by integration of (17) subject to boundary condition (10)

$$w=\int_r^1 r\,W\,dr \tag{18}$$

With $\Phi$ and $W$ expressions (2) and (8) for stress components and for stress resultants become

$$\sigma_r=\Phi \qquad \sigma_\theta=r\frac{d\Phi}{dr}+\Phi \tag{19}$$

$$\left.\begin{aligned} &M_r=-\left(\frac{dW}{dr}+\left(1+\nu\right)W\right) \qquad M_\theta=-\left(\nu\frac{dW}{dr}+\left(1+\nu\right)W\right) \\ &Q_r=-\left(r\frac{d^2W}{dr^2}+3\frac{dW}{dr}\right) \end{aligned}\right\} \tag{20}$$

Equations (16) are of second order so we need four boundary conditions. At the plate edge, the two conditions follow from the pressure boundary condition (9) and plate support boundary conditions (11).

$$\Phi\left(1\right)=-p \tag{21}$$

$$\left.\begin{aligned} &W\left(1\right)=0 \qquad \left(\text{clamped edge}\right) \\ &\left[\frac{dW}{dr}+\left(1+\nu\right)W\right]\left(1\right)=0 \qquad \left(\text{hinged edge}\right) \end{aligned}\right\} \tag{22}$$

At the plate center, we have, from (13) and (14), $\sigma_r(0) = \sigma_\theta(0)$ and $M_\theta(0) = M_r(0)$. So that expressions (19) for membrane stresses and (20) for stress resultants fulfill these two conditions, we must set

$$\frac{d\Phi}{dr}(0) = 0 \qquad \frac{dW}{dr}(0) = 0 \tag{23}$$

Using L'Hospital rule and the boundary conditions (12), we find that $W(0) = \dfrac{d^2 w}{dr^2}(0)$ and $\Phi(0) = \dfrac{d^2\phi}{dr^2}(0)$. Hence, from (15), we have

$$\Phi(0) \neq \infty \qquad W(0) \neq \infty \tag{24}$$

We can equally well use either of conditions (23) or (24).

## 3 Buckling pressure

If we discard quadratic terms in the first equation of the system (16) we obtain well-known decupled system

$$\left.\begin{aligned} \frac{d^2\Phi}{dr^2} + \frac{3}{r}\frac{d\Phi}{dr} = 0 \\ \frac{d^2W}{dr^2} + \frac{3}{r}\frac{dW}{dr} - \Phi W = 0 \end{aligned}\right\} \tag{25}$$

The solution of this system, which satisfy boundary conditions (21) and (24), is

$$\Phi = -p \tag{26}$$

$$W = C\frac{J_1\left(\sqrt{p}r\right)}{r} \tag{27}$$

where $J_n(.)$ is the $n$th order Bessel function of first kind and $C$ is an arbitrary constant. Introducing this into expressions (19) for the membrane stresses and (18) for the deflection, we found

$$\sigma_r = \sigma_\theta = -p \tag{28}$$

$$w = \frac{C}{\sqrt{p}}\left[1 - J_0\left(\sqrt{p}r\right)\right] \tag{29}$$

Substituting (27) into the boundary condition (22) leads to the following characteristic equations for the buckling (critical) pressure $p_c$

$$\left.\begin{aligned} J_1\left(\sqrt{p_c}\right) = 0 \quad & \left(\text{clamped edge}\right) \\ \sqrt{p_c}J_0\left(\sqrt{p_c}\right) - \left(1-\nu\right)J_1\left(\sqrt{p_c}\right) = 0 \quad & \left(\text{hinged edge}\right) \end{aligned}\right\} \tag{30}$$

Either of these equations has an infinite number of roots. For the clamped edge, the first root is well-known critical pressure $p_c$ [6, 23]

$$p_{c,1} = 14.68197... \tag{31}$$

The higher critical pressures are: $p_{c,2} = 49.218456...,\ p_{c,2} = 103.499450...$

Unlike clamped edge, the characteristic equation for hinged edge contains Poisson ratio $\nu$. To obtain dependence of critical pressure $p_c$ on $\nu$ we assume expansion in the form

$$p_{c,1}\left(\nu\right) = p_0 + p_1\nu + p_2\nu^2 + \cdots$$

Substituting this into boundary condition (30) for hinged edge and then equating like powers $\nu$ we as constant term obtain the transcendental equation for $p_0$

$$\sqrt{p_0}J_0\left(\sqrt{p_0}\right) - J_1\left(\sqrt{p_0}\right) = 0 \tag{32}$$

while for $p_1, p_2, \ldots$ the equations are linear. The numerical calculation gives the following result

$$p_{c,1}\left(\nu\right) = 3.389958 + 2.836835\,\nu - 0.496654\,\nu^2 + 0.055319\,\nu^3 + \cdots \tag{33}$$

For the range $0 < \nu \le 0.5$ this formula is correct to at least three decimal places. For example, for $\nu = 0.5$ the first zero of (30) is $p_{c,1} = 4.690998$ while the above formula gives $p_{c,1} = 4.691$. For $\nu = 0.3$ we obtain $p_{c,1} = 4.1978$ which is the value that matches with those given in [24]. Similarly, we obtain

$$p_{c,2}\left(\nu\right) = 28.424282 + 2.072928\,\nu - 0.002756\,\nu^2 - 0.025054\,\nu^3 + \cdots \tag{34}$$

etc.

## 4 Integration by power series

The power series in $r$ for unknowns $\Phi$ and $W$, which are consistent with the form of the equations system (25) and also satisfy boundary conditions (24) and (23), contains only even powers of $r$ [19]

$$\Phi = \sum_{n=0}^{\infty} \Phi_n r^{2n} \tag{35}$$

$$W = \sum_{n=0}^{\infty} W_n r^{2n} \tag{36}$$

Using these the expressions (18) for the plate deflection, (19) for the membrane stresses, and (20) for the stress resultants become

$$w = -\tfrac{1}{2} \sum_{n=0}^{\infty} \frac{W_n}{n+1} \left(1 - r^{2n+2}\right) \tag{37}$$

$$\sigma_r = \sum_{n=0}^{\infty} \Phi_n r^{2n} \qquad \sigma_\theta = \sum_{n=0}^{\infty} \left(2n+1\right) \Phi_n r^{2n} \tag{38}$$

$$\left.\begin{aligned} M_r &= -\sum_{n=0}^{\infty} \left(2n+1+\nu\right) W_n r^{2n} \qquad M_\theta = -\sum_{n=0}^{\infty} \left[\nu\left(2n+1\right)+1\right] W_n r^{2n} \\ Q_r &= -4 \sum_{n=0}^{\infty} \left(n^2-1\right) W_n r^{2n+1} \end{aligned}\right\} \tag{39}$$

$$\frac{dw}{dr} = \sum_{n=0}^{\infty} W_n r^{2n+1} \tag{40}$$

To obtain coefficients $\Phi_n$ and $W_n$ we introduce (35) and (36) into (16). Arranging and then equating coefficients of corresponding powers of $r$ leads to the following recursive system of equations

$$\left.\begin{aligned} \Phi_n &= -\frac{1}{8n\left(n+1\right)} \sum_{k=0}^{n-1} W_k W_{n-k-1} \\ W_n &= \frac{1}{4n\left(n+1\right)} \sum_{k=0}^{n-1} W_k \Phi_{n-k-1} \qquad \left(n = 1,2,\ldots\right) \end{aligned}\right\} \tag{41}$$

Once we knew $W_0$ and $\Phi_0$ we can using these formulas successively compute all the coefficients of the series (35)-(36).

Inspection of (41) reviles that the explicit dependence of coefficients $\Phi_n$ and $W_n$ on $\Phi_0$ and $W_0$ has the following form

$$\left.\begin{aligned} \Phi_n &= \sum_{k=0}^{\lfloor (n-1)/2 \rfloor} A_{n,k} W_0^{2k+2} \Phi_0^{n-2k-1} \\ W_n &= \sum_{k=0}^{\lfloor n/2 \rfloor} B_{n,k} W_0^{2k+1} \Phi_0^{n-2k} \qquad (n = 1,2,3,\ldots) \end{aligned}\right\} \tag{42}$$

where $A_{n,k}$ on $B_{n,k}$ are coefficients depends only on $n$ and $k$., $\lfloor . \rfloor$ is the floor function. These formulas are useless for theoretical consideration because explicit formulas for $A_{n,k}$ and $B_{n,k}$ are not known. However, $A_{n,k}$ and $B_{n,k}$ can be calculated numerically so the above formulas can be useful for numerical calculation of $W_0$ and $\Phi_0$.

*Calculation of the coefficients.* Now, by substituting expansion (35) for $\Phi$ into pressure boundary condition (21) we obtain

$$\sum_{n=0}^{\infty} \Phi_n = -p \tag{43}$$

Expressing $\Phi_n$ by (42) yield implicit relationship between $\Phi_0$ and $W_0$

$$F\left(\Phi_0, W_0\right) = -p \tag{44}$$

Introducing (36) for $W$ into the plate support boundary condition (22) yield

$$\left.\begin{aligned} \sum_{n=0}^{\infty} W_n &= 0 \qquad (\text{clamped edge}) \\ \sum_{n=0}^{\infty} \left(2n + 1 + \nu\right) W_n &= 0 \qquad (\text{hinged edge}) \end{aligned}\right\} \tag{45}$$

Substituting $W_n$ from (42) into above leads to yet another implicit relationship between $\Phi_0$ and $W_0$

$$G\left(\Phi_0, W_0\right) = 0 \tag{46}$$

This relationship together with (44) forms system of two equations for unknowns $\Phi_0$ and $W_0$. If $p > p_{c,1}$ is given a then the system (44)-(46) can be solved numerically for a finite number of $\Phi_n$ and $W_n$. However, because we want to construct various diagrams which shows the dependence of the plate variables on *p*, such approach is not convenient. Therefore we for the calculation of coefficient suggest the following procedure:

**For** $W_0 \in \left(0, W_0^{(c)}\right)$

Solve $G\left(\Phi_0, W_0\right) = 0 \quad \Rightarrow \quad \Phi_0$

Calculate $p = -F\left(\Phi_0, W_0\right)$

Calculate coefficients $\Phi_n$ and $W_n$ using (41)

Calculate variables of interest using (37), (38), (39)

**end**

The only problem remains a numerical solution of the equation $G\left(\Phi_0, W_0\right) = 0$ for the unknown $\Phi_0$; this must be done by some care because when $W_0$ is given then the equation has two solutions. This is a consequence of the of the presence of turning point (Fig 2 and Fig 3) i.e. point where $dW_0 / d\Phi_0 = 0$. However, the coordinates of a turning point $\Phi_0^{(c)}$ and $W_0^{(c)}$ can be calculated in advance by solving simultaneous equations

$$G = 0 \qquad \frac{\partial G}{\partial \Phi_0} = 0 \tag{47}$$

In the practical calculation, one should scan values of $W_0$ from zero up to turning point and then back. The numerical experiments show that for calculation of unknown $\Phi_0$ with the precision of say six decimal places, we need up to about a hundred terms in the series (35)-(36) (Table 1, Table 2, Table 3).

*Numerical results.* With the proposed method, we calculate the variation of radial membrane stresses at the plate centers and the variation of deflection of the plate at the center with boundary pressure (Fig 3 and Fig 4). The graphs match with those given by Friedrichs and Stoker [5] and Bodner [6]. However, the present diagrams cover a bit larger range of the pressure; up to $p / p_c = 21$ for both kinds of plate support compare to $p/p_c = 15$ for hinged plate in [5] and $p/p_c = 5$ for the clamped plate in [6]. Agreement between results from [5] and results obtained by the present method can also be seen from the Table 4 where some numerical values are given.

*Convergence.* Study of the convergence of the series solution (35)-(36) is beyond the scope of this article (see [7] for more details about convergence investigation). However, some numerical experiments indicate that the power series method works well only for the first branch of possible solutions of the problem (Fig 1 and Fig 2). Numerical calculations show that for other solution branches convergence is nor achieved even with thousands of coefficients (up to 20000) in quad precision arithmetic.

At the end of this section, we note that the proposed method of calculation of the coefficients differs from one suggested by Friedrichs and Stoker [4-6, 19]. Namely, they purpose a method where $\Phi_0$ and $W_0$ are to be chosen and then a nonlinear equation resulting from boundary condition is to be solved for the unknown auxiliary variable. As they observed, the method needs a good estimate for $\Phi_0$ and $W_0$ which is not an easy task.

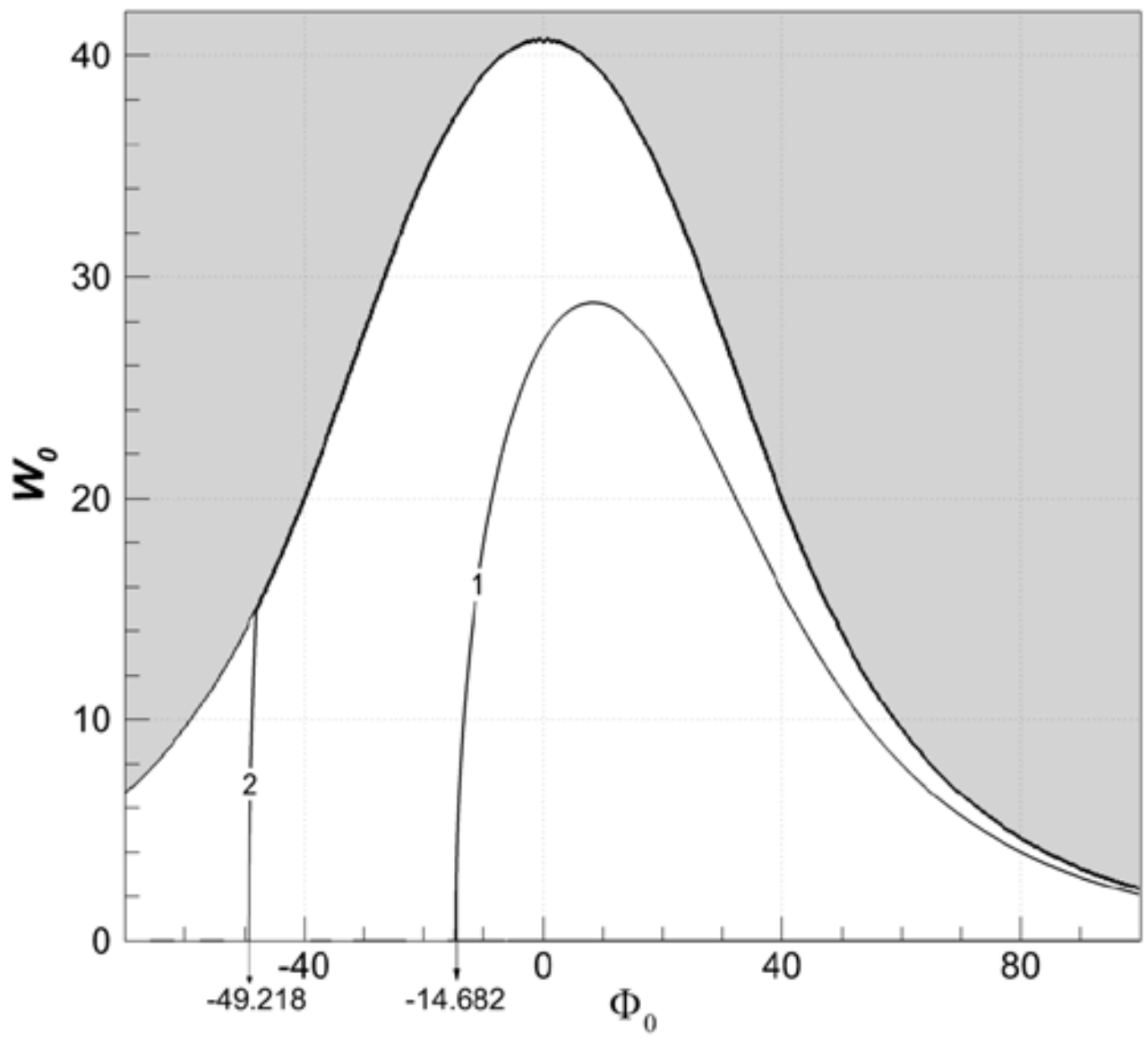

**Figure 1.** Graph of the function $G\left(\Phi_0, W_0\right) = 0$ for the case of clamped plate. On shaded region, series $G\left(\Phi_0, W_0\right) = 0$ fails to converge. The turning point of the first branch is at $\Phi_0 = 8.358896$ and $W_0 = 28.851493$ . $N \leq 1183$ coefficients was used to construct the graph where the test of convergence was $\left|G_N - G_{N-1}\right| < 10^{-7}$ and divergence with $\left|G_N - G_{N-1}\right| > 10^4$ for ten occurrences.

**Table 1.** Convergence study. Coordinates of the turning of point of $G\left(\Phi_0^c, W_0^c\right) = 0$ for the clamped plate. $N$ is number of terms in series (42) for $\Phi_n$ and $W_n$ . Numerical precision for calculation was set to 16 digits.

| $N$ | $\Phi_0^{(c)}$ | $W_0^{(c)}$ | $p/p_c$ | $\sigma_r(0)/p_c$ | $w(0)$ |
|---|---|---|---|---|---|
| 32 | 8.**108315** | 28.8**23866** | 2.**799231** | -0.**197291** | -10.**813531** |
| 64 | 8.358896 | 28.85149**4** | 2.832**777** | -0.2009**80** | -10.912**035** |
| 128 | 8.358896 | 28.851493 | 2.832764 | -0.200978 | -10.911995 |

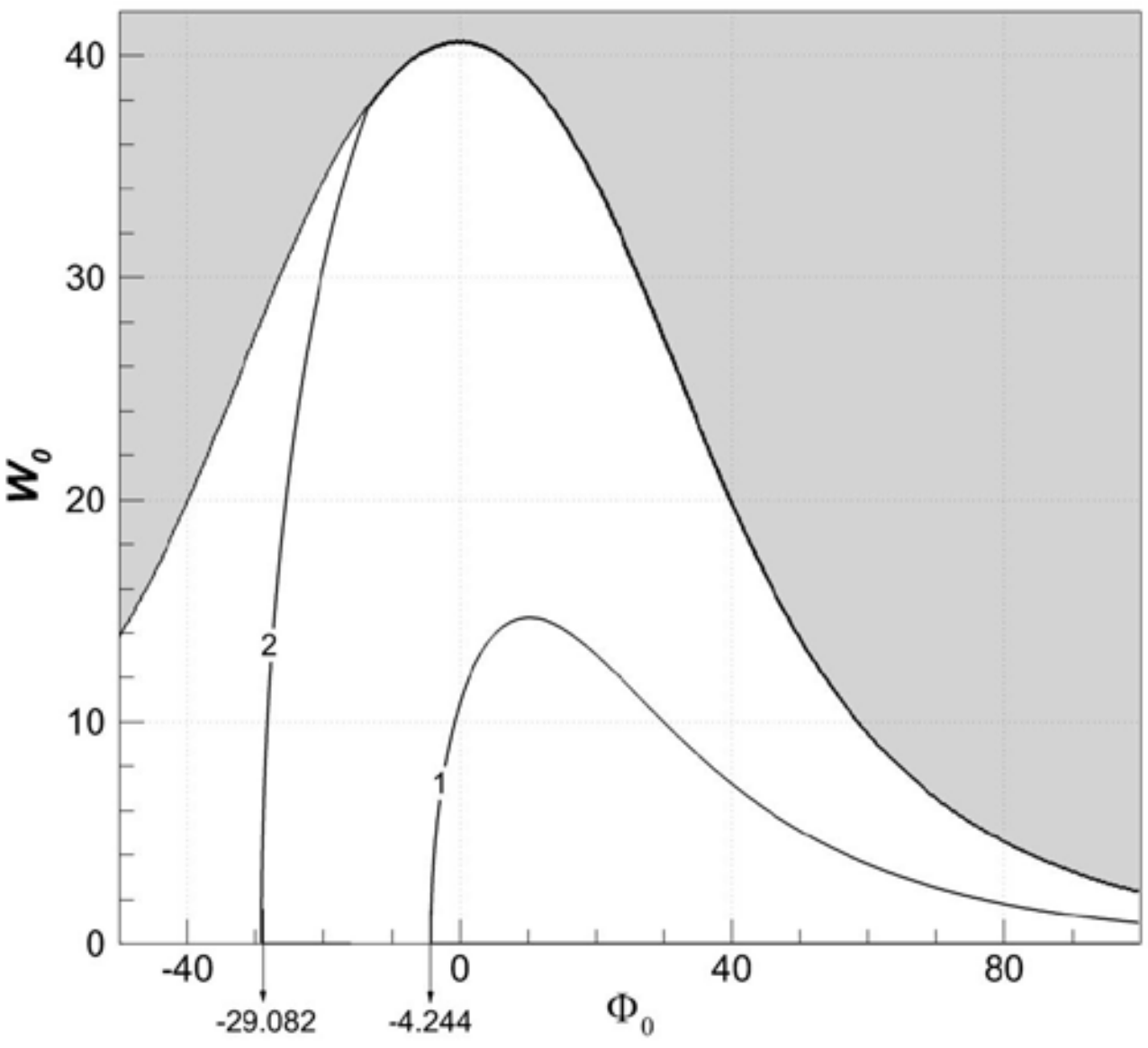

**Figure 2.** Graph of $G\left(\Phi_0, W_0\right) = 0$ for the case of the hinged plate when $\nu = 0.318$. On shaded region, series $G\left(\Phi_0, W_0\right) = 0$ fails to converge. The turning point of the first branch is at $\Phi_0 = 10.204570$ and $W_0 = 14.681282$ . $N \leq 2551$ coefficients was used to construct the graph where the test of convergence was $\left|G_N - G_{N-1}\right| < 10^{-20}$ and divergence with $\left|G_N - G_{N-1}\right| > 10^6$ for ten occurrence.

**Table 2.** Convergence study. Coordinates of the turning of point of $G\left(\Phi_0^c, W_0^c\right) = 0$ for the hinged plate with $\nu = 0.318$. $N$ is number of terms in series (42) for $\Phi_n$ and $W_n$ . Numerical precision for calculation was set to 16 digits.

| $N$ | $\Phi_0^{(c)}$ | $W_0^{(c)}$ | $p/p_c$ | $\sigma_r(0)/p_c$ | $w(0)$ |
|---|---|---|---|---|---|
| 16 | 10.20**9110** | 14.6**79893** | 3.4**80020** | -0.691**307** | -10.57**8707** |
| 32 | 10.204570 | 14.681**028** | 3.4795**70** | -0.6910**70** | -10.576784 |
| 64 | 10.204570 | 14.681282 | 3.479568 | -0.691090 | -10.576784 |

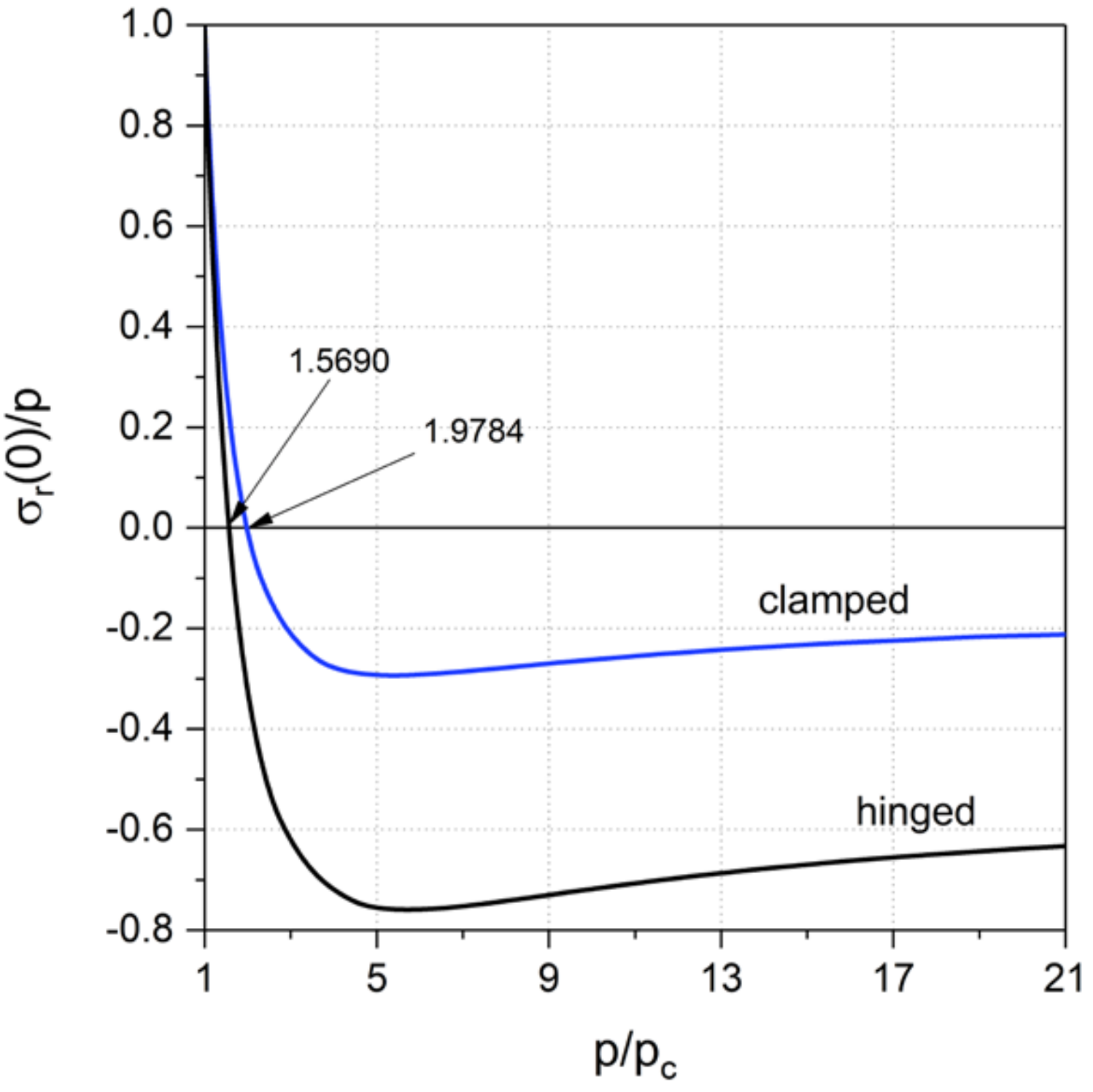

**Figure 3.** Radial membrane stress at plate center versus normalized edge pressure. $\nu = 0.318$ .

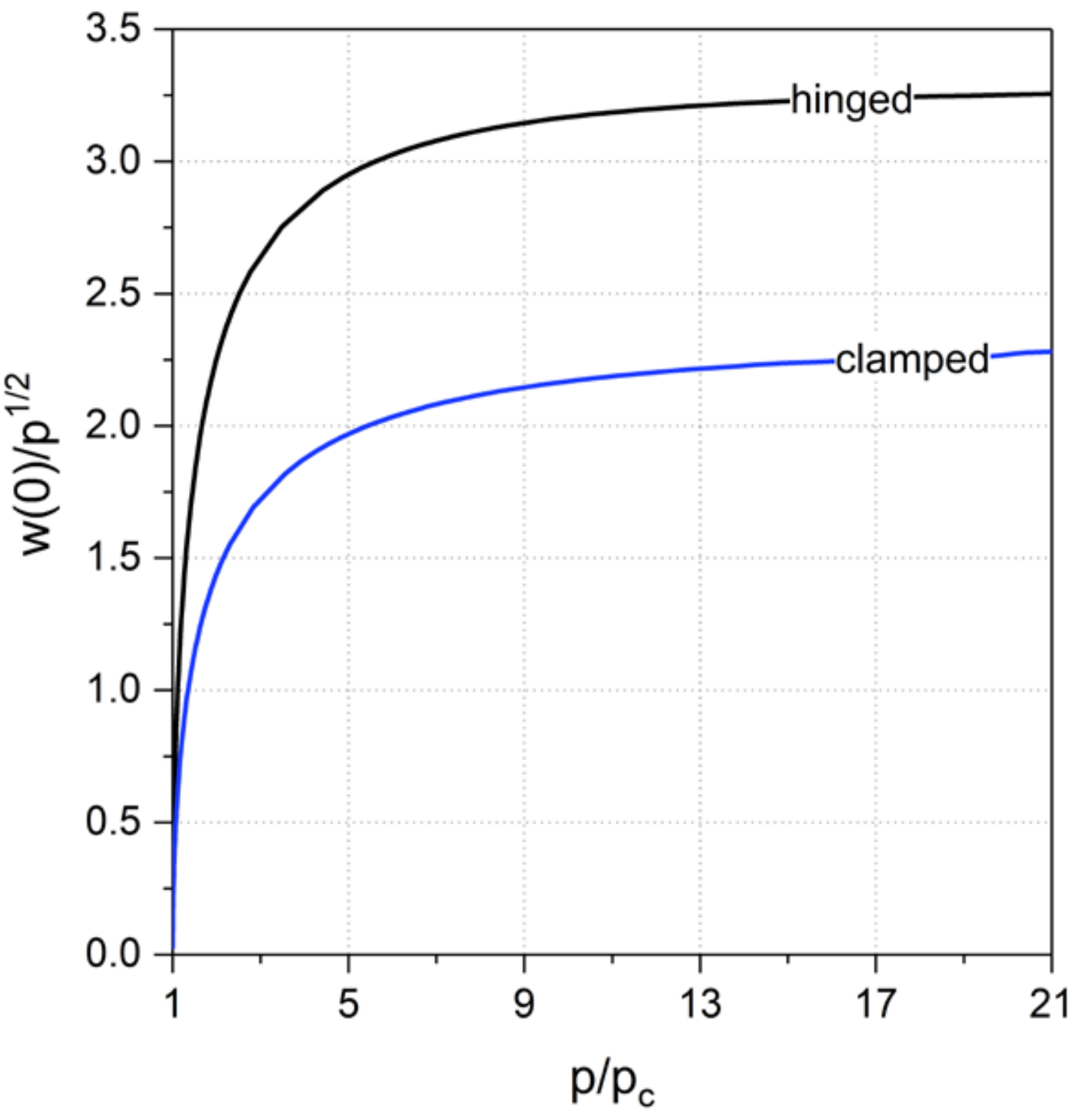

**Figure 4.** Plate deflection at center versus normalized edge pressure. $\nu = 0.318$

**Table 3.** Convergence study for the hinged plate. Values of stress and deflection at plate center for $p/p_c = 15$ , $\nu = 0.318$ . $N$ is number of series terms.

| | $N$ | | | |
|---|---|---|---|---|
| | 32 | 64 | 128 | collocation |
| $\sigma_r/p(0)$ | 0.6699 **4057** | 0.6698 357**3** | 0.6698 3575 | 0.6698 3575 |
| $w/\sqrt{p}(0)$ | 3.228**5 0663** | 3.2284 396**4** | 3.2284 3967 | 3.2284 397 |

**Table 4.** Comparison of results for the hinged plate. $\nu = 0.318$

| $\frac{p}{p_c}$ | $\sigma_r(0)/p$ | | | $w(0)/\sqrt{p/p_c}$ | | |
|---|---|---|---|---|---|---|
| | Ref [5] | series | collocation | Ref [5] | series | collocation |
| 1.065 | 0.823 | 0.8223 | 0.822314 | 1.58 | 1.5798 | 1.580012 |
| 2.01 | -0.326 | -0.3290 | -0.329007 | 4.65 | 4.6555 | 4.655455 |
| 14.74 | -0.673 | -0.6719 | -0.671887 | 6.65 | 6.6473 | 6.646723 |

## 5 Numerical integration

An alternative to the series solution is a numerical solution. For that, we use collocation code *colnew*. [18]. To use the code we introduce the following new unknown functions

$$z_1 \equiv \Phi \quad z_2 \equiv \frac{d\Phi}{dr} \quad z_3 \equiv W \quad z_4 \equiv \frac{dW}{dr} \quad z_5 \equiv w \tag{48}$$

Note that among unknowns we also include the deflection $w$. With these the system of equations (16) together with $(17)_2$ become

$$z_1'' = -\frac{3}{r} z_2 + \frac{1}{2} z_3^2 \qquad z_3'' = -\frac{3}{r} z_4 + z_1 z_3 \qquad z_5' = r z_3 \tag{49}$$

The boundary conditions for this system are, by (23),(10) and (21),(22),

$$z_2(0) = 0 \quad z_4(0) = 0 \tag{50}$$

$$\begin{gathered} z_5(1) = 0 \quad z_1(1) = -p \\ \left.\begin{array}{ll} z_3(1) = 0 & (\text{clamped edge}) \\ z_4(1) + (1+\nu) z_3(1) = 0 & (\text{hinged edge}) \end{array}\right\} \end{gathered} \tag{51}$$

The problem is highly nonlinear. Therefore, to assure the convergence, we start the calculation with a pressure, which is a bit above the critical and at each iteration the pressure is gradually increased until it reaches given pressure $p$. For an initial guess of the solution, we use analytical expressions (28) and (29). In all calculations, we set absolute error tolerance to $10^{-7}$ and use four collocation points per interval.

We use the described procedure to calculate the distribution of displacement, stress components along plate radius (Fig 4-9). Distributions of the plate variables shown on Figs 7-9 for hinged plate match those given in [5]. Numerical values given in Table 5-6 shows that the collocation method can handle the pressure up to $10^4$. For higher pressures, the calculation becomes time-consuming.

With the collocation method, we can also easily handle higher buckling modes. The graphs in Fig 10 and Fig 11 depicts the deflection of the plate center versus applied pressure for different buckling modes. These graphs show the well-known fact, that the plate under increasing pressure cannot jump to higher order symmetric buckled mode. More on secondary buckling see [15-17, 20]… On Figs 12-13 the examples of higher buckled plate shapes are show. We note that these forms are unstable.

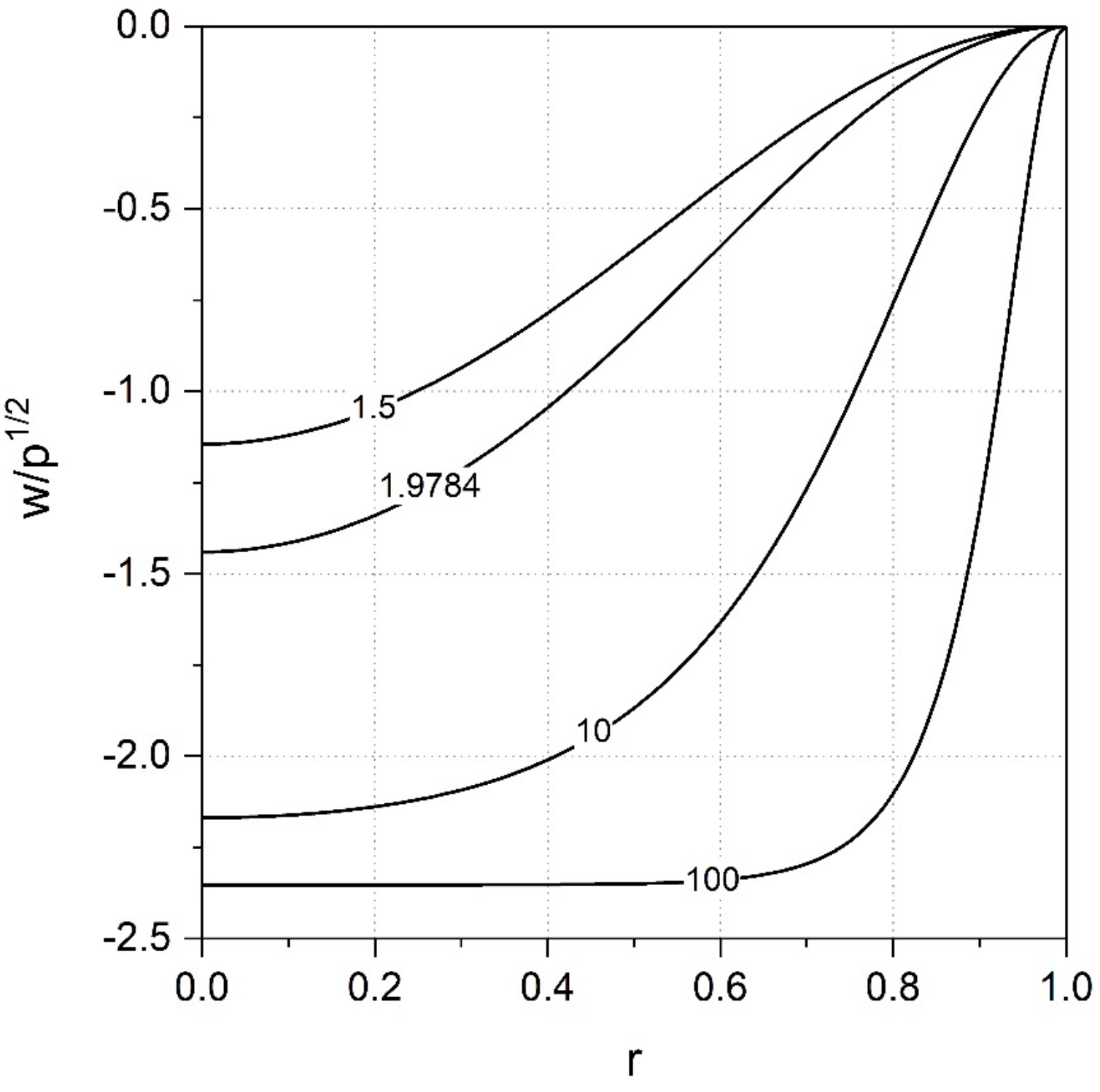

**Figure 4.** Clamped plate. Deflection for different value of $p/p_c$

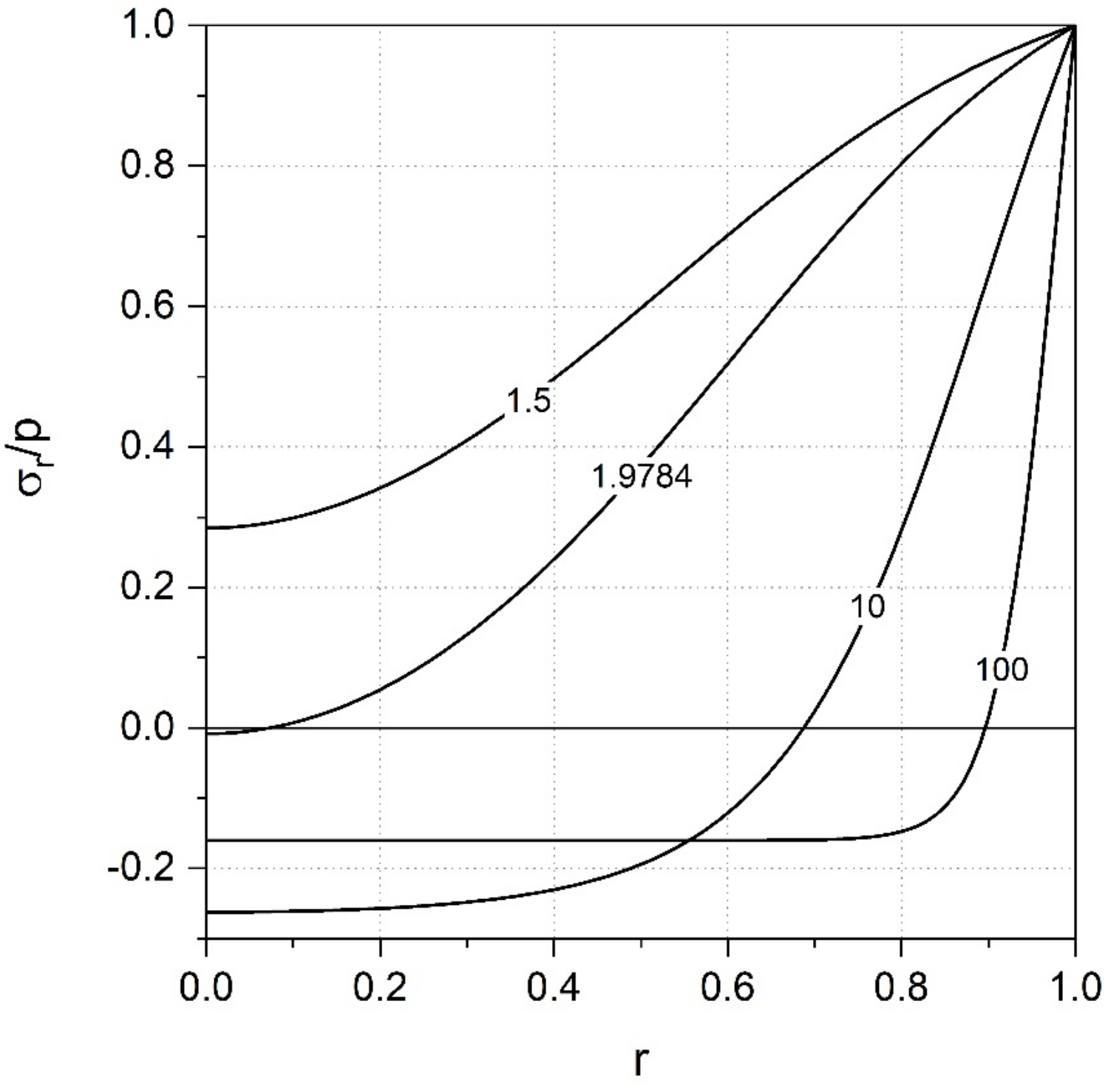

**Figure 5.** Clamped plate. Radial membrane stresses for different value of $p/p_c$

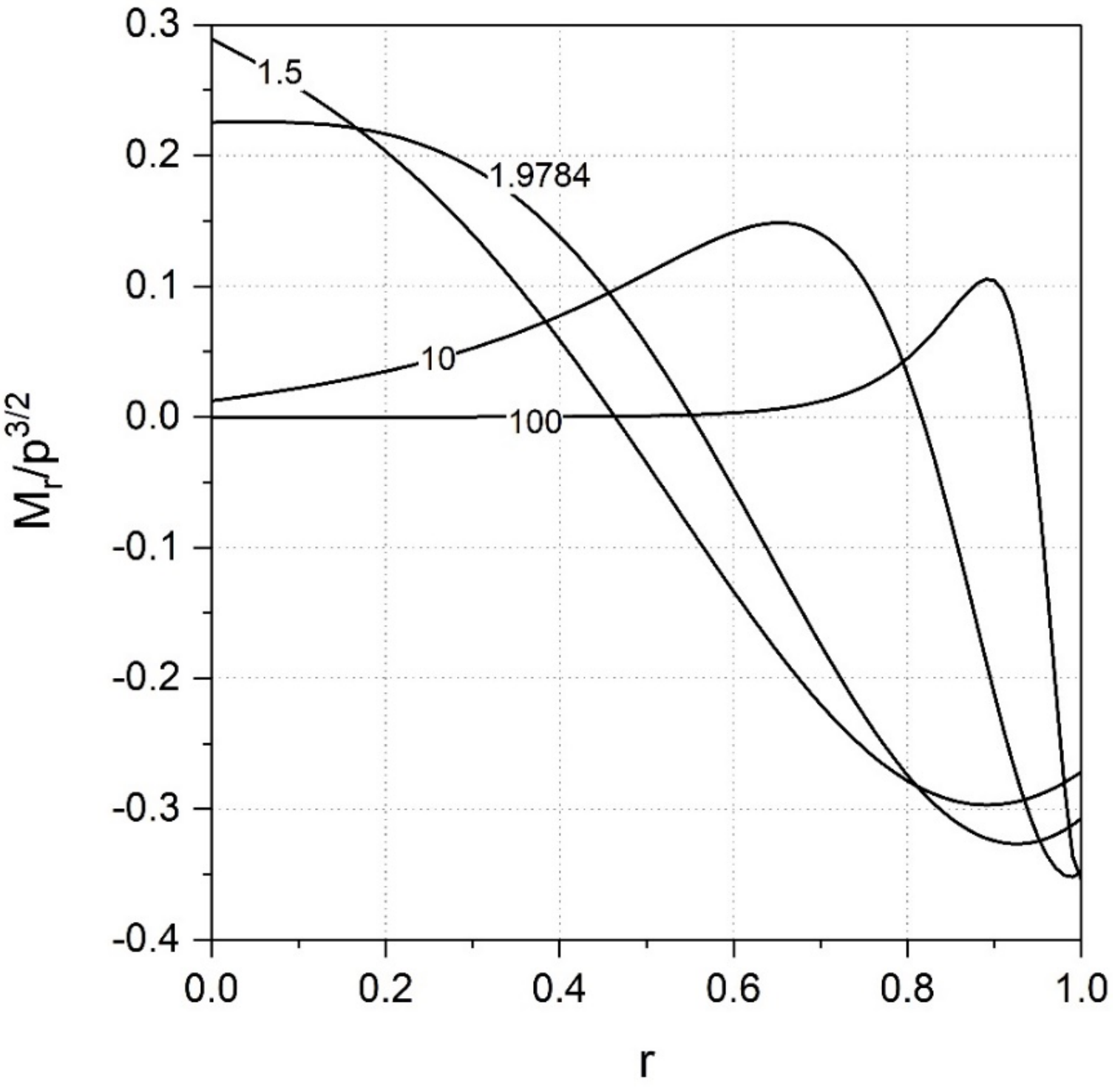

**Figure 6.** Clamped plate. Radial bending moment for different value of $p/p_c$

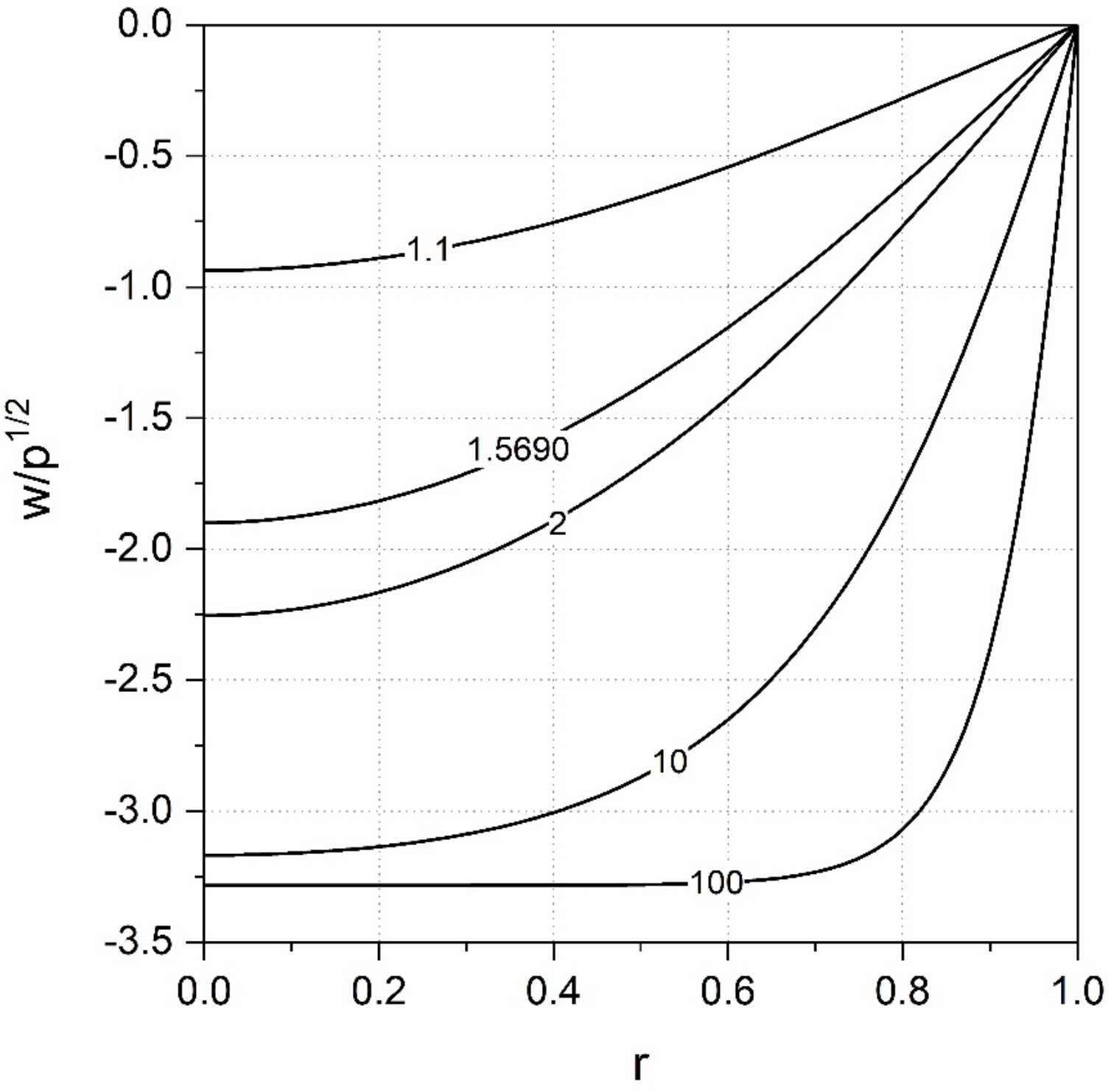

**Figure 7.** Hinged plate. Deflection for different value of $p/p_c$

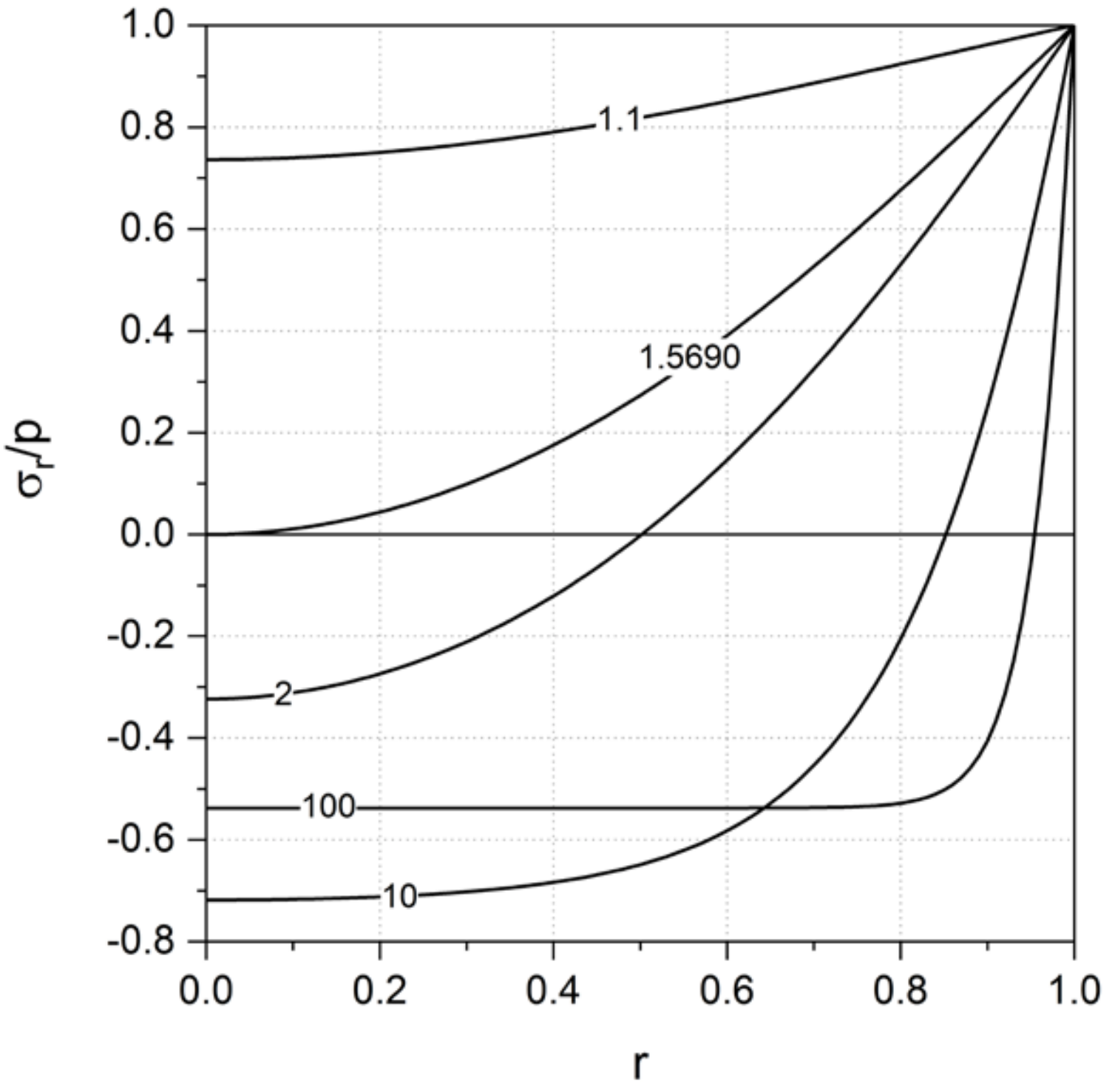

**Figure 8.** Hinged plate. Radial membrane stresses for different value of $p/p_c$

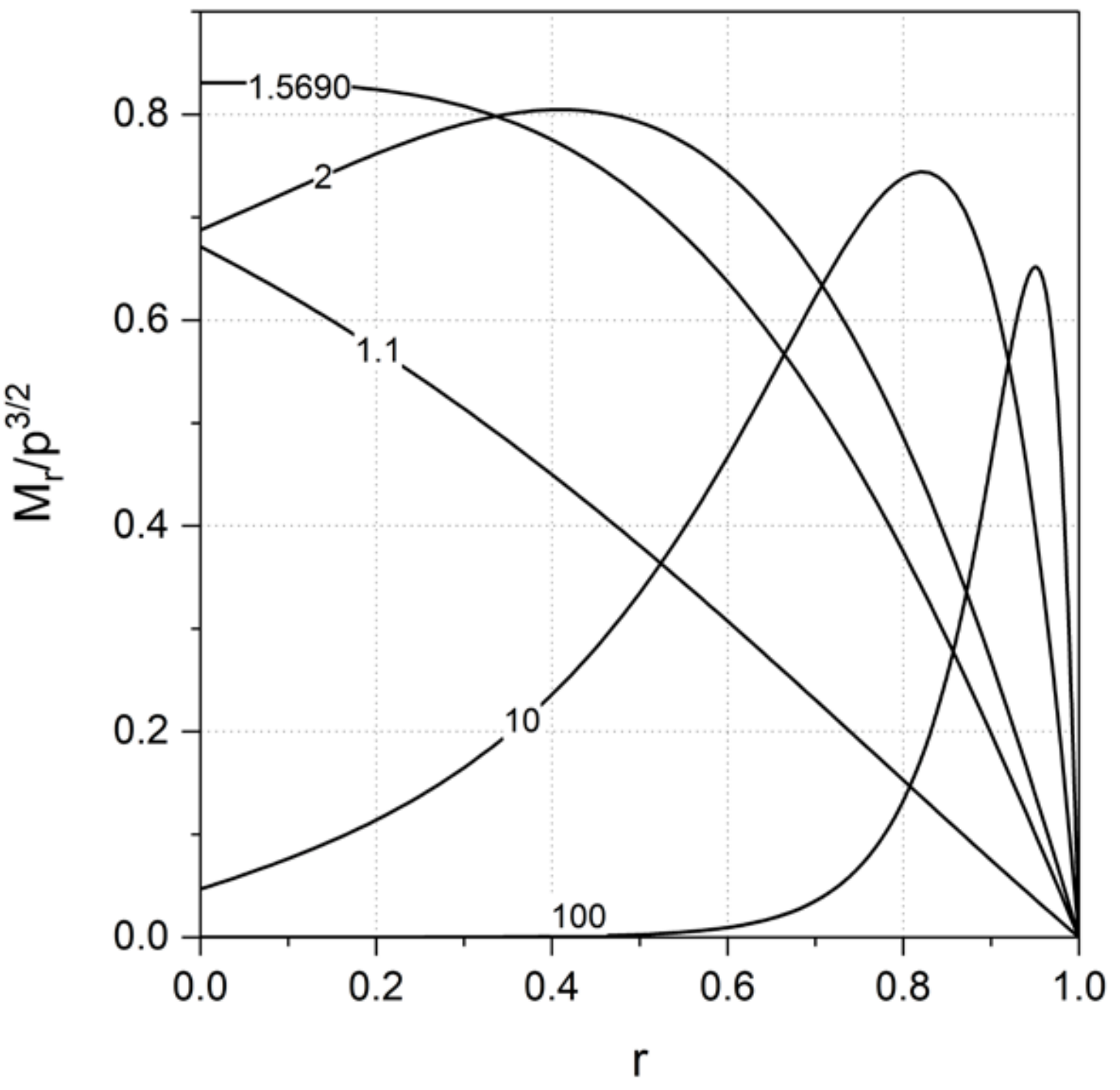

**Figure 9.** Hinged plate. Radial bending moment for different value of $p/p_c$

**Table 5.** Values calculated by collocation method for clamped plate. Grayed numbers are boundary condition.

| $p/p_c$ | $r$ | $\sigma_r/p$ | $\sigma_\theta/p$ | $M_r/p\sqrt{p}$ | $M_\theta/p\sqrt{p}$ | $w/\sqrt{p}$ |
|---|---|---|---|---|---|---|
| 1.5 | 0 | 0.284560 | 0.060637 | 0.289136 | 0.289136 | -1.144848 |
| 10 | 0 | -0.262430 | -0.021658 | 0.012292 | 0.012292 | -2.168812 |
| 100 | 0 | -0.160446 | -0.004187 | 0.000001 | 0.000001 | -2.352612 |
| 1000 | 0 | -0.138598 | -0.001144 | 0.000000 | 0.000000 | -2.391103 |
| 10000 | 0 | -0.132516 | -0.000346 | 0.000000 | 0.000000 | -2.403225 |
| 1.5 | 1 | 1.000000 | 0.304357 | -0.271564 | -0.086357 | 0.000000 |
| 10 | 1 | 1.000000 | 0.341294 | -0.348426 | -0.110799 | 0.000000 |
| 100 | 1 | 1.000000 | 0.350268 | -0.354553 | -0.112748 | 0.000000 |
| 1000 | 1 | 1.000000 | 0.353332 | -0.354890 | -0.112855 | 0.000000 |
| 10000 | 1 | 1.000000 | 0.354331 | -0.354844 | -0.112840 | 0.000000 |

**Table 6.** Values for hinged plate. Grayed numbers are boundary condition.

| $p/p_c$ | $r$ | $\sigma_r/p$ | $\sigma_\theta/p$ | $M_r/p\sqrt{p}$ | $M_\theta/p\sqrt{p}$ | $w/\sqrt{p}$ |
|---|---|---|---|---|---|---|
| 1.1 | 0 | 0.7363949 | 0.3408371 | 0.6714738 | 0.6714738 | -0.9372574 |
| 10 | 0 | -0.7184642 | -0.1102904 | 0.0469087 | 0.0469087 | -3.1681835 |
| 100 | 0 | -0.5376272 | -0.0260984 | 0.0000037 | 0.0000037 | -3.2833401 |
| 1000 | 0 | -0.4920994 | -0.0075541 | 0.0000000 | 0.0000000 | -3.2903261 |
| 10000 | 0 | -0.4787423 | -0.0023240 | 0.0000000 | 0.0000000 | -3.2923844 |
| 1.1 | 1 | 1.0000000 | 0.6349260 | 0.0000000 | 0.2675114 | 0.0000000 |
| 10 | 1 | 1.0000000 | 1.5263145 | 0.0000000 | 0.2166180 | 0.0000000 |
| 100 | 1 | 1.0000000 | 1.5979634 | 0.0000000 | 0.0701980 | 0.0000000 |
| 1000 | 1 | 1.0000000 | 1.6102070 | 0.0000000 | 0.0222623 | 0.0000000 |
| 10000 | 1 | 1.0000000 | 1.6131383 | 0.0000000 | 0.0070432 | 0.0000000 |

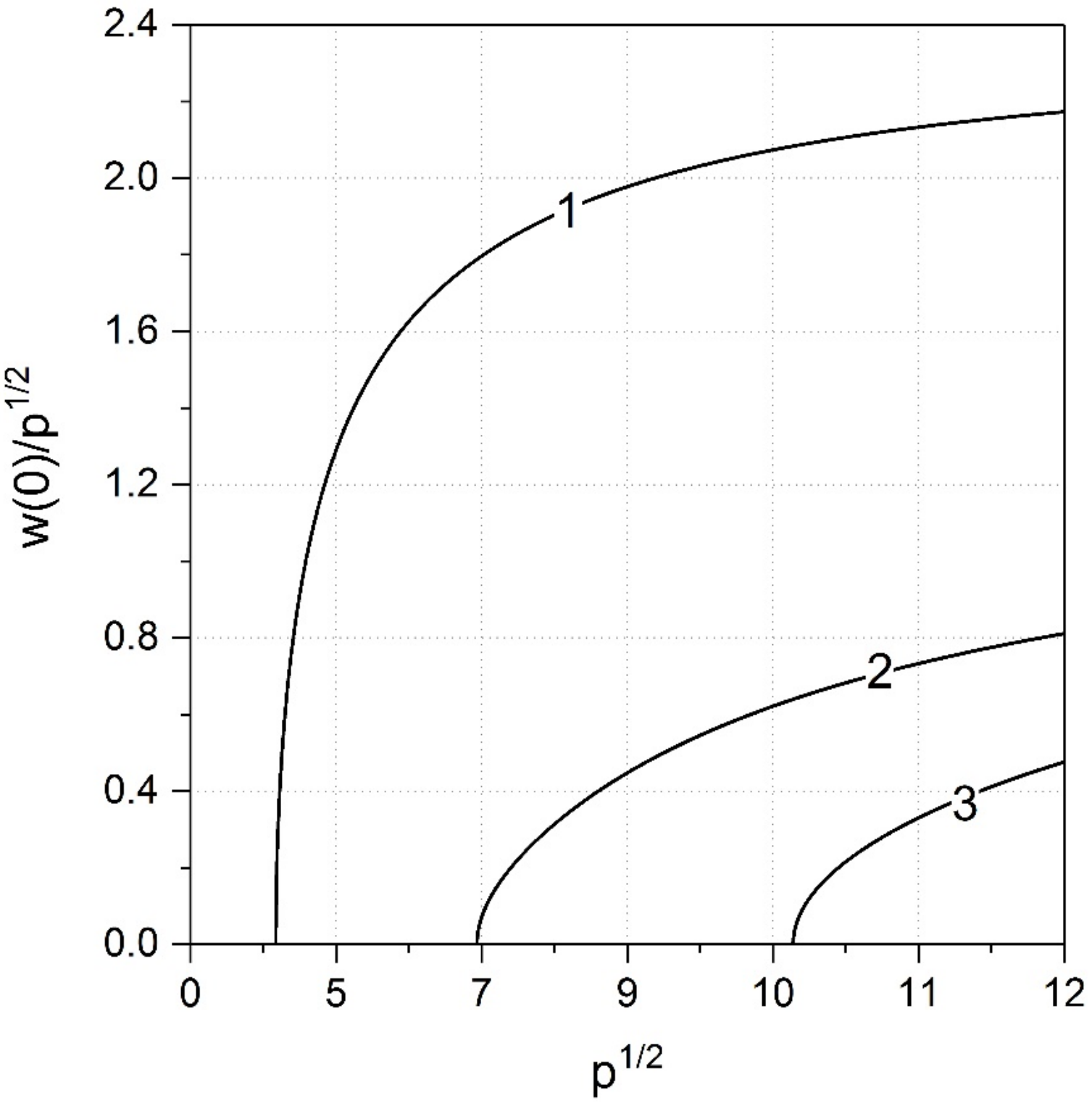

**Figure 10.** Clamped plate. Deflection at center versus pressure for different buckling modes.

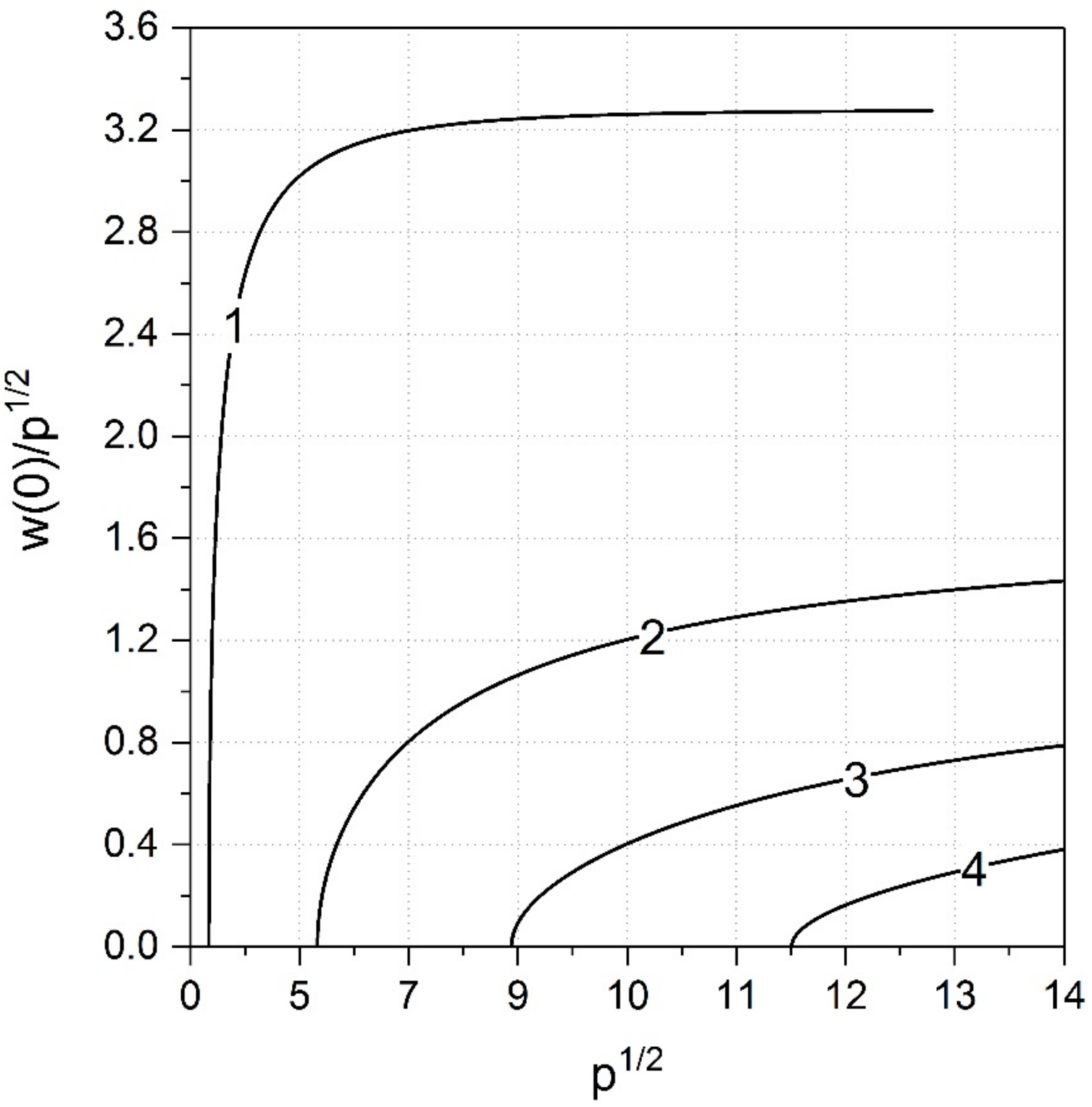

**Figure 11.** Hinged plate. Plate deflection at center versus pressure for different buckling modes.

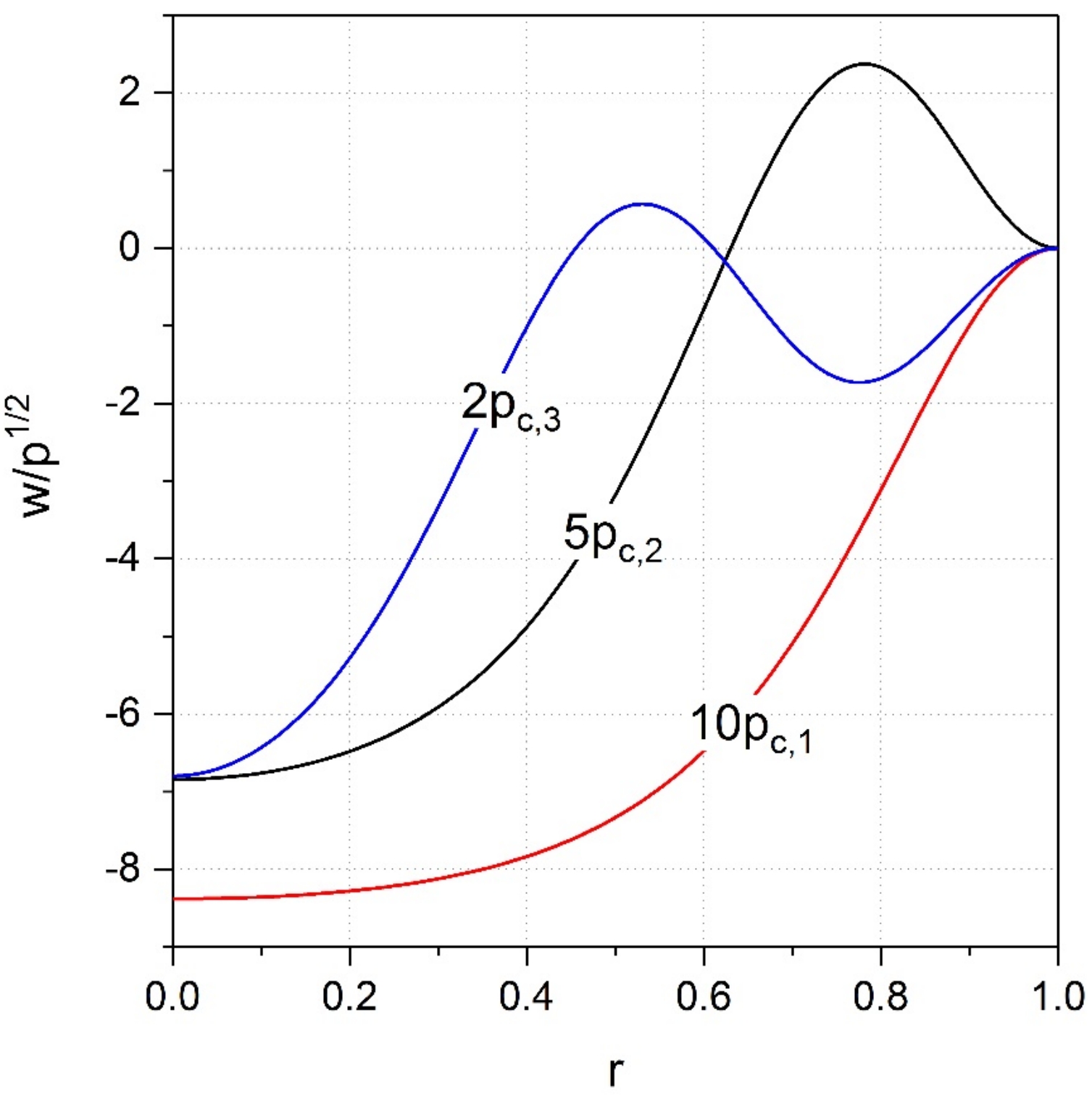

**Figure 12.** Deflection of clamped plate for various values of pressure. $p_{c,n}$ is the critical pressure for $n$th buckled mode.

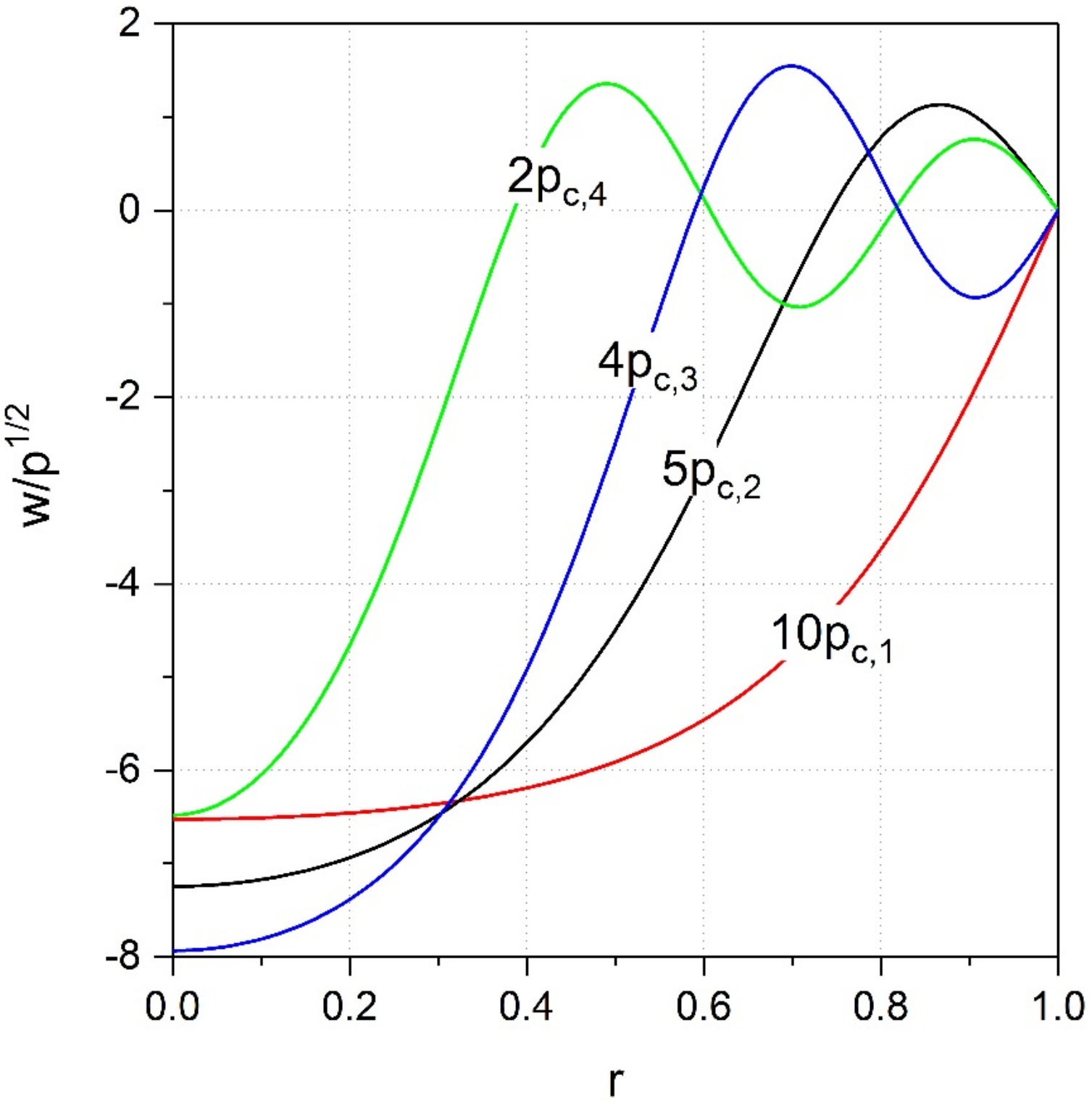

**Figure 13.** Deflection of a hinged plate for various values of pressure. $p_{c,n}$ is the critical pressure for $n$th buckled mode.

## Conclusions

In the article we for the case of hinged plate give an explicit formula for critical pressure depends on Poisson's ratio. Also, we describe an alternative method for calculation of coefficients of series solution which does not require an accurate initial guess. By numerical calculations, we show that the series solution is useful only for the first buckling mode. A reason for this is an open question. Lastly, we demonstrate that the collocation method works well with high edge pressure and can also be used to calculate a higher buckling mode of the plate.